\RequirePackage{letltxmacro}
\LetLtxMacro{\LaTeXtextbf}{\textbf}

\documentclass{ieeeaccess}
\LetLtxMacro{\textbf}{\LaTeXtextbf}

\usepackage{cite}
\usepackage{amsmath,amssymb,amsfonts}
\usepackage{algorithmic}
\usepackage{graphicx}
\usepackage{textcomp}

\usepackage[utf8]{inputenc} 
\usepackage[T1]{fontenc}    
\usepackage{hyperref}       
\usepackage{url}            
\usepackage{booktabs}       
\usepackage{amsfonts}       
\usepackage{nicefrac}       
\usepackage{microtype}      
\usepackage{lipsum}
\usepackage{fancyhdr}       
\usepackage{graphicx}       
\graphicspath{{media/}}     

\usepackage{multirow}
\usepackage{float}
\usepackage{subfig} 
\usepackage{amsmath}
\usepackage{mathtools}
\usepackage{bm}
\usepackage{multirow}
\usepackage{enumitem}
\usepackage[ruled,vlined,linesnumbered,noend]{algorithm2e}
\usepackage[table,xcdraw]{xcolor}

\usepackage{combelow}

\usepackage{xcolor}

\def\BibTeX{{\rm B\kern-.05em{\sc i\kern-.025em b}\kern-.08em
    T\kern-.1667em\lower.7ex\hbox{E}\kern-.125emX}}
\begin{document}
\history{Date of publication 8 November 2023, date of current version 14 November 2023.}
\doi{10.1109/ACCESS.2023.3331220}

\title{MCWDST: a Minimum-Cost Weighted Directed Spanning Tree Algorithm for Real-Time Fake News Mitigation in Social Media}
\author{
\uppercase{Ciprian-Octavian Truic\u{a}}\authorrefmark{1,*},
\uppercase{Elena-Simona Apostol}\authorrefmark{1,*},
\uppercase{Radu-C\u{a}t\u{a}lin Nicolescu}\authorrefmark{1,*}, and 
\uppercase{Panagiotis Karras}\authorrefmark{2}
}

\address[1]{Computer Science and Engineering Department, Faculty of Automatic Control and Computers, National University of Science and Technology Politehnica Bucharest, 060042 Bucharest, Romania}
\address[2]{Department of Computer Science, University of Copenhagen, 2100 Copenhagen, Denmark}
\address[*]{These authors contributed equally to this work}

\tfootnote{The publication of this paper is supported by the National University of Science and Technology Politehnica Bucharest through the PubArt program.}

\markboth
{C.O. Truic\u{a} \headeretal: MCWDST: a Minimum-Cost Weighted Directed Spanning Tree Algorithm for Real-Time Fake News Mitigation in Social Media}
{C.O. Truic\u{a} \headeretal: MCWDST: a Minimum-Cost Weighted Directed Spanning Tree Algorithm for Real-Time Fake News Mitigation in Social Media}

\corresp{Corresponding author: E.S. Apostol (e-mail: elena.apostol@upb.ro).}

\begin{abstract}
The widespread availability of internet access and handheld devices confers to social media a power similar to the one newspapers used to have. People seek affordable information on social media and can reach it within seconds. Yet this convenience comes with dangers; any user may freely post whatever they please and the content can stay online for a long period, regardless of its truthfulness. A need arises to detect untruthful information, also known as \emph{fake news}. In this paper, we present an end-to-end solution that accurately detects fake news and immunizes network nodes that spread them in real-time. To detect fake news, we propose two new stack deep learning architectures that utilize convolutional and bidirectional LSTM layers. To mitigate the spread of fake news, we propose a real-time network-aware strategy that (1) constructs a minimum-cost weighted directed spanning tree for a detected node, and (2) immunizes nodes in that tree by scoring their harmfulness using a novel ranking function. We demonstrate the effectiveness of our solution on five real-world datasets.
\end{abstract}


\titlepgskip=-21pt

\maketitle

\section{Introduction}\label{sec:introduction}

With the accelerated technology adoption by a growing number of users, social media have become the main medium for the dissemination of information on current news and events~\cite{Truica2021}. While these new media bring several benefits (e.g., a large number of consumers reached, instant and continuous updates on one's topics of interest), they also enable the spread of harmful information in the form of fake news, and may thus polarize public discourse regarding critical topics (e.g., elections~\cite{Pyrhonen2019}, vaccination~\cite{Petit2021}, health hazards~\cite{Naeem2020}) and threaten democratic values~\cite{Rose2019}. Because of its detrimental effects on society at large~\cite{truica2023danes,apostol2023contcommrtd,coban2023contain}, the \emph{fake news} phenomenon has been studied by scientists and practitioners alike; fake news is defined as news articles that intentionally contain verifiably false misleading information inconsistent with factual reality ~\cite{Allcott2017,Molina2021,Wang2020,Gelfert2018,brody2018model,Ilie2021,Truica2022}. To mitigate the threat of fake news, journalists have started to manually classify news and offer websites with fact-checking mechanisms that provide a verdict regarding its veracity, such as \href{https://www.politifact.com/}{PolitiFact} and \href{https://www.snopes.com/}{Snopes}. However, such solutions may fail in high-velocity information spreading social media, as news appears and spreads much faster than any manual verification; by the time it is checked, the news may have been already shared with many sources and its negative effect may have taken hold. In this paper, we propose new models and strategies for misinformation detection and mitigation to address the current real-world challenges posed by fake news.

In particular, we aim to:
\begin{itemize}
    \item[$(O_1)$] Propose new deep learning architectures to accurately detect fake news in social media; and
    \item[$(O_2)$] Propose new real-time strategies to mitigate the spread of detected fake news in a social network.
\end{itemize}

To reach these objectives, we answer the following questions:
\begin{itemize}
    \item[$(Q_1)$] Can we improve the accuracy of fake news detection using new deep learning architectures? 
    \item[$(Q_2)$] Can we immunize nodes that spread harmful content using network information in real-time?
\end{itemize}

To answer $(Q_1)$ and achieve objective $(O_1)$, we propose two novel stack deep learning architectures that utilize convolutional and bidirectional LSTM layers. We show the effectiveness of these architectures with ample benchmarking on four real-world datasets. To answer $(Q_2)$ and reach objective $(O_2)$, we propose a real-time algorithm for network immunization, which builds a minimum-cost weighted directed spanning tree for a detected source node $r$ and chooses nodes in that tree to immunize by scoring their potential harmfulness using a ranking function. The proposed ranking functions consider the following network information for each node:
\begin{itemize}
     \item[(1)] How long a chain of followers is (i.e., the length of the diffusion path), 
    \item[(2)] How many nodes it reaches (i.e., the spread of information), and 
    \item[(3)] How fast it spreads information (i.e., the information diffusion speed).
\end{itemize}

To show the effectiveness of network immunization, we evaluate our method on one real-world Twitter dataset. In summary, the main contributions of this paper are:
\begin{itemize}
    \item[(1)] New deep learning architectures for fake news detection;
    \item[(2)] New real-time fake news mitigation strategy consisting of an algorithm for building the minimum-cost weighted directed spanning tree for a given node and a network-aware node harmfulness ranking function;
    \item[(3)] Benchmarking on multiple real-world datasets to evaluate the efficiency of our deep learning architectures for fake news detection; 
    \item[(4)] Evaluation of our real-time mitigation algorithm on a real-world Twitter dataset.
\end{itemize}

The rest of this paper is structured as follows: In Section~\ref{sec:stateoftheart}, we present the state of the art. In Section~\ref{sec:metodology}, we present the proposed architecture, deep learning models for fake news detection, minimum-cost weighted directed spanning tree algorithm for a given source node and harmful node ranking function. In Section~\ref{sec:implementation}, we detail the implementation of our architecture. In Section~\ref{sec:results}, we describe the datasets and analyze in detail the experimental validation of our solution. Section~\ref{sec:discussions} provides an in-depth discussion of results, and Section~\ref{sec:conclusions} concludes the paper and hints at future research.

\section{Related Work}\label{sec:stateoftheart}

In this section, we present the approaches of previous research on the two tasks of interest to our work: detecting fake news and mitigating its spread.

\subsection{Fake news detection}

Nakamura et al.~\cite{Nakamura2020Fakeddit} introduced a dataset comprising \href{https://www.reddit.com/}{Reddit} posts along with an architecture designed to identify posts that contain fake pieces of information using image data, title data, or both combined. The advantages of working with such a dataset are its diversity, large size, and multidimensionality, which allows researchers to treat fake news detection as a 2-way, 3-way, or 6-way classification problem. Further, the authors combine text and image features to classify whether a post spreads fake pieces of information or not. Text is embedded using the BERT~\cite{devlin2019bert} and InferSent~\cite{conneau2017supervised} models, while image data is extracted using the VGG16~\cite{simonyan2015deep}, ResNet50~\cite{he2016deep}, and EfficientNet~\cite{tan2019efficientnet} models.

Kumar et al.~\cite{Kumar2020} tackle the fake news detection challenge via sentiment analysis, implementing seven deep learning architectures, such as long short-term memory (LSTM), bidirectional LSTM, convolutional neural networks (CNN), and ensemble models that combine the aforementioned. 
After performing fake news classification using these models, the authors evaluate the models' performance against classic machine learning algorithms, such as logistic regression and support vector machines.

Khan et al.~\cite{khan2021benchmark} address the fake news detection task via classical machine learning algorithms and more advanced deep-learning and neural network models, and compare their performance on two well-known datasets: Liar and Fake or Real News, along with a self-built dataset, which the authors claim to be more variate and denser than the other two.

Granik and Mesyura~\cite{Granik2017} tackle the fake news detection problem by using the Naive Bayes classification algorithm, which, despite its simplicity, yields good results on the task. Moreover, the authors suggest further adjustments that can be made to improve the results of the aforementioned method.

P{\'e}rez-Rosas et al.~\cite{perezrosas2018automatic} collect two original fake news detection datasets and use a linear kernel SVM algorithm to perform classification. The first dataset is built by collecting real news from reliable US news websites and adding pieces of fake information to them to turn them into fake news. For the second dataset, the authors target celebrity news and search the web for pairs of fake and real articles regarding celebrity gossip. The classification model takes as input a set of predefined linguistic features or a mix of them, such as unigrams or bigrams, punctuation marks, psycholinguistic features, readability, and syntax. The authors evaluate the model's performances using these features on the two created datasets.

Nguyen et al.~\cite{nguyen-etal-2019-fake} suggest that implicit correspondences between articles can help improve the performance of a fake news detection algorithm. The authors convert the task of detecting whether an article contains untruthful pieces of information into an inference problem in a Markov random field and transform the algorithm that solves this kind of problem into a deep neural network.

Truic{\u{a}} et al.~\cite{Truica2022b} propose the use of transformer-based sentence embeddings and transfer learning for fake news detection for news articles in both English and German.

In their study, Truic{\u{a}} and Apostol~\cite{Truica2023} show empirically that the embedding used for encoding the data is one of the key factors in accurately determining the veracity of news articles. 

Mayank et al~\cite{mayank2021deapfaked} propose DEAP-FAKED, a new model that uses Natural Language Processing techniques, Graph Neural Networks, and Knowledge graphs to identify Fake News. The experimental results on the \href{https://www.kaggle.com/c/fake-news/}{Kaggle} dataset
show this approach improves the performance of misinformation detection. We use the DEAP-FAKED model in our comparison.

Truic{\u{a}} and Apostol~\cite{Truica2022} designed MisRoB{\AE}RTa, a transformed-based ensemble model used for multi-class fake news detection. The ensemble uses RoBERTa~\cite{Liu2019} and BART~\cite{Lewis2020} to encode the textual content of the news. The experimental results on the Kaggle and The \href{https://github.com/several27/FakeNewsCorpus}{Fake News Corpus (FNC)}  datasets show how this approach improves the overall performance on the task of misinformation detection. We use MisRoB{\AE}RTa to compare the results obtained by our proposed models on Kaggles.

Raza et al.~\cite{Raza2022} propose FND-NS, a framework that learns useful representations for predicting fake news. FND-NS uses both the textual content and the social context to determine the veracity of news articles and social media posts. For the Fakeedit corpus, we compare the results of our models with the ones obtained by FND-NS.

\subsection{Fake news mitigation}

Sharma et al.~\cite{sharma2021network} start out with graphs constructed from two Twitter datasets and design two independent diffusion paths for real and fake posts, each with a separate set of learned parameters. Using these paths and the obtained parameters, they determine the user characteristics that matter most in news spread and propose two methods to mitigate the spread of fake news: blocking a node (i.e., the user sending the news) or blocking the edge (i.e., the news transmission path). Their experiments indicate that an account that spreads fake news has a low follower count and no account description available, while an account that posts real news has a large number of followers, and is associated with well-known sources.

Saxena et al.~\cite{saxena2020competitive} consider a social network as a graph in which there are three types of users: positive --- which spread true news, negative --- which spread false news, and neutral --- which are influenced by the other two types. The authors model these influences probabilistically, where neutral nodes influenced by one of the two types of news are less likely to change stance, the longer the diffusion process lasts. To mitigate false information circulating through the graph, the authors design an algorithm that assigns each true news broadcaster a metric called ``truth score'', and selects the top-$k$ sources with the best scores.

Shu et al.~\cite{shu2020fakenewsnet} contribute to the field of false news detection and mitigation by building a comprehensive dataset. They collect news from \href{https://www.politifact.com/}{PolitiFact}, \href{https://www.gossipcop.com/}{GossipCop}, and \href{https://www.eonline.com/}{E! Online}.
The first two are platforms where the level of truth of news is labeled following a fact-checking process, while the latter is considered a credible news source. The dataset includes user features, i.e., number of followers, number of followees, user location (if mentioned in the profile description), user comments or retweets, as well as post features, i.e., post content (written or visual). These features are obtained by searching the news collected from the 3 sources using the Twitter search tool. By virtue of its diverse features, the dataset can be used in both fake news detection and mitigation tasks. Their best-performing model on this dataset is CNN~\cite{shu2020fakenewsnet}, which we will also use in our comparison.

Sayyadiharikandeh et al.~\cite{Sayyadiharikandeh_2020} address the problem of fake news mitigation assuming that non-human entities in social networks are set up to spread false information in an automated way. The authors implement a mechanism to detect such entities, which they call Botometer. They build several models to detect people and different types of bots in a social network. The models are aggregated into an ensemble learning algorithm, which produces the final classification results.

Nevertheless, none of the above approaches assembles a combined fake news detection and mitigation pipeline, as we do, further discussed in the following section.
Firstly, we identify the harmful nodes that spread fake news in the network using a novel deep learning architecture.
Secondly, we detect and rank harmful nodes using a novel algorithm that uses the direct weighted graph structure.
Lastly, we immunize the network using a blocking strategy based on the ranked list of harmful nodes.

\section{Methodology}\label{sec:metodology}

\begin{figure}[!htbp]
\centering
\includegraphics[width=1\columnwidth]{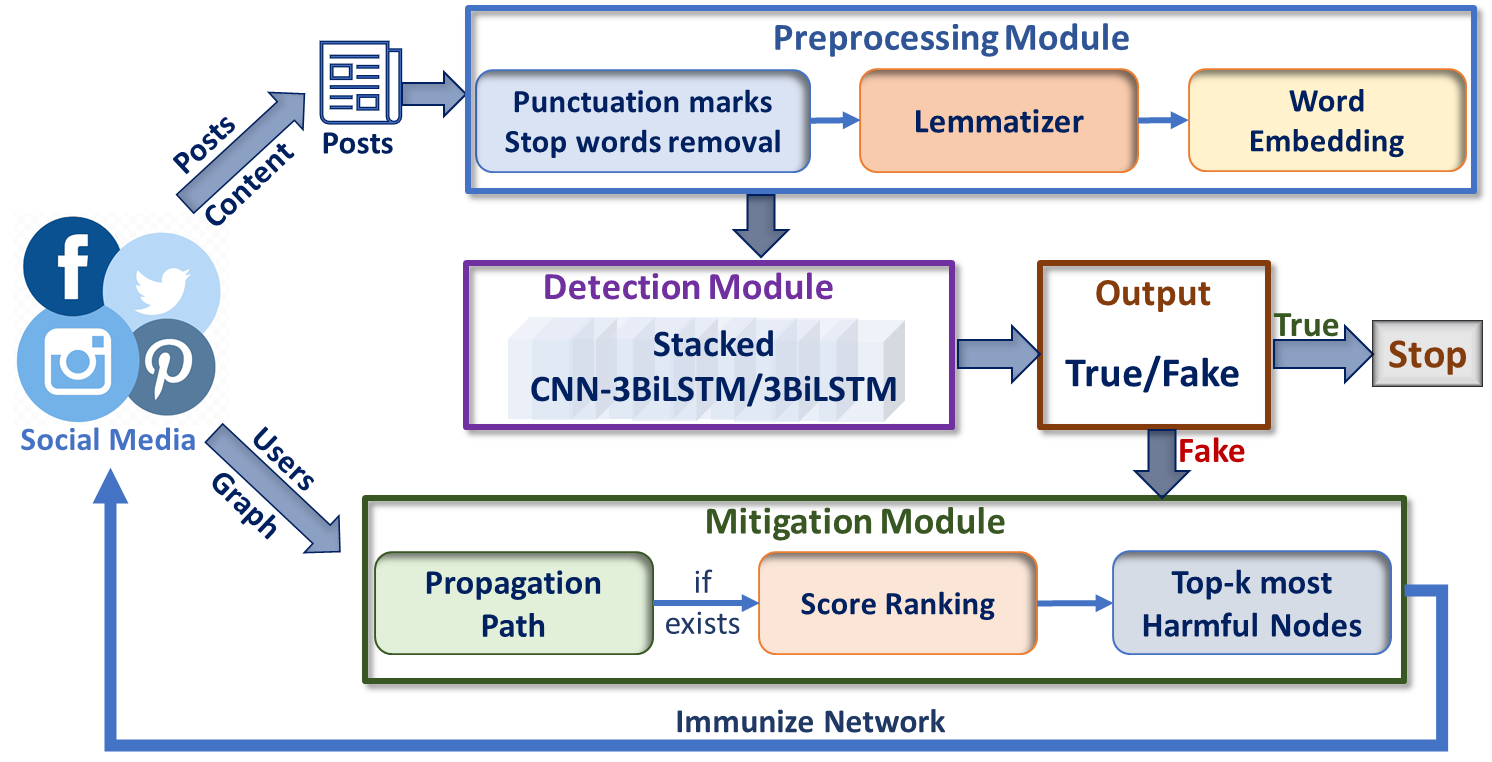}
\caption{Detection and Mitigation Pipeline}
\label{fig:pipeline}
\end{figure}

Figure~\ref{fig:pipeline} presents our pipeline for fake news detection and mitigation. We mine a Social Media platform in real-time to detect nodes that spread harmful content; for each such node, we immunize the network with a real-time mitigation strategy. To be labeled as true or fake, a post passes through a preprocessing stage, where its content is edited and converted to a real-number matrix, which is passed to the detection stage. The detection stage is represented by one of our two proposed architectures, which outputs a post label. In case the post is false, we start the mitigation process; we construct a propagation path, i.e., a minimum-cost weighted directed spanning tree starting from each node that spreads fake news, compute a score for and rank the harmfulness of each node, and eventually extract the top-$k$ most harmful nodes.

\subsection{Preprocessing}

The preprocessing module cleans the text and minimizes the vocabulary while preserving meaning; it comprises three steps:
(1) punctuation marks and stop words removal,
(2) lemmatization, and
(3) word encoding using word embeddings.

Due to their high frequency of occurrence and specificity, stop words and symbols do not add relevant information when fed into the detection model. Thus, they may better be removed from posts. In the first stage of preprocessing, we remove these two types of elements. In the second stage, lemmatization, we transform words with similar forms into one unique form to homogenize the content of the posts. For the posts to be included in our proposed models, each word is transformed into a numerical representation using a word embedding vector. We use two pre-trained word embedding models and a model we trained on the chosen datasets.

\subsection{Fake news detection}

The purpose of classification is to predict previously unseen items based on inferences derived by training on a set comprising news articles/posts and their labels. We first describe the basic elements of the two proposed architectures and motivate why we chose them.

We use an \textit{Embedding} layer at the start of our models to store word embeddings that we obtain through the continuous bag-of-words model or pre-trained models. Thereby, our architectures have access to a word embedding dictionary. In general, an embedding layer would either generate embeddings or map every word of a post to a real-number vector representation; since we generate word embeddings in the preprocessing stage, we use the embedding layer with the latter function.

We employ a \textit{Bidirectional Long Short-Term Memory (BiLSTM)} layer because the news we predict represents word sequences, and this type of layer is specialized in learning long-term dependencies in text. A further reason to employ such a layer is that it uses two simple LSTM architectures that look both forward and backward in the sequence, such that given a sequence element, both previous and future elements are available to the network.

We use a \textit{Convolutional} layer with $N$ filters and kernel size $k$ to extract patterns from a $k$-size window in the data passed to the network, creating $N$ features by means of a convolution operation between the text window and every distinct filter, and adding a bias term. Every filter has an associated channel where it stores features. The final result of the Convolutional layer with $N$ filters and size $k$ applied on a post of length $L$ consists of $N$ channels of length $L-k+1$, each containing the new features obtained by its corresponding filter. We equalize post lengths in the preprocessing stage by zero-padding or truncation, such that the post length is constant, leading to constant channel size for all posts. Following the Convolutional layer, we use a \emph{max-pooling} layer with pooling size $p$ to decrease the size of the feature channels by grouping elements into groups of length $p$ and choosing only the feature with the maximum value. We use \emph{dropout} layers to deactivate a percentage of the outputs coming from the previous layer; thus, we reduce overfitting by creating artificial noise and improve the generalization capacity of the network when new, unknown data is fed for prediction. We use \emph{dense} layers with linear activation as a connection between the network layers, and a final dense layer with a softmax activation function to produce the classification result, i.e., the probability of the post to be true.

Concretely, we propose two deep learning models, namely CNN-3BiLSTM and 3BiLSTM. The main difference between them is that CNN-3BiLSTM creates new features through a convolutional layer, from which it extracts the best-generated features using a max-pooling layer, while 3BiLSTM feeds the initial features directly to the BiLSTM and dropout sequences.

\begin{figure}[!htbp]
\centering
\includegraphics[width=1\columnwidth]{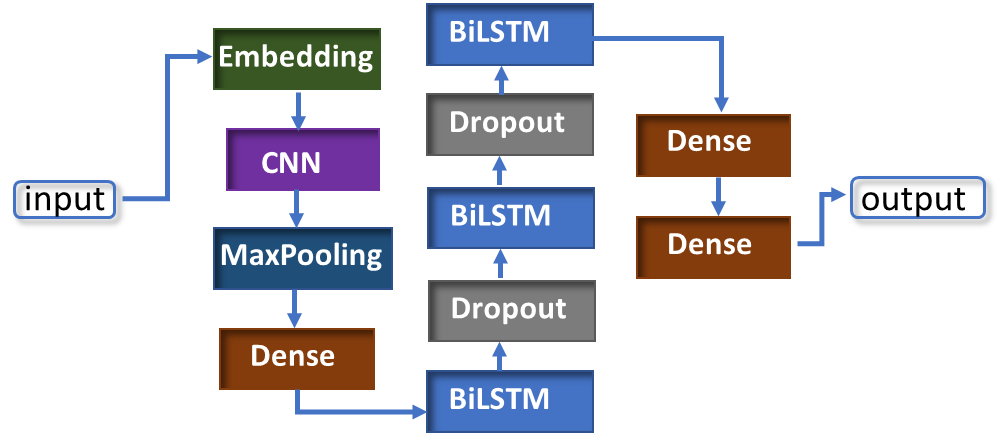}
\caption{Stacked CNN-3BiLSTM architecture}
\label{fig:arch-conv}
\end{figure}

In the stacked CNN-3BiLSTM architecture (Figure~\ref{fig:arch-conv}), we use an \emph{embedding} layer where we pass word embeddings as weight parameters, followed by a \emph{convolutional} layer with one group of 128 filters and kernel size 3 with ReLU as activation function and a \emph{max-pooling} layer with pool size 2 and the number of strides set to 2. We add a \emph{dense} layer with 256 units, followed by three BiLSTM layers having dropout layers in-between. We set the number of units for the BiLSTM layers to 64 and the dropout rate to 0.2. The architecture ends with two dense layers: one with 128 linear activation units and one with 1 sigmoid activation unit.

\begin{figure}[!htbp]
\centering
\includegraphics[width=1\columnwidth]{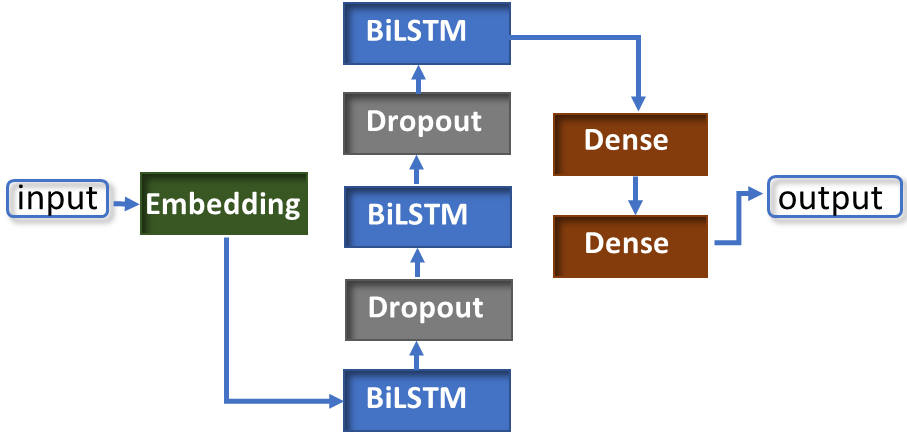}
\caption{Stacked 3BiLSTM architecture}
\label{fig:arch-stack}
\end{figure}

In the stacked 3BiLSTM architecture (Figure~\ref{fig:arch-stack}), we use an embedding layer with weights set as the word embeddings, then stack three 128 units BiLSTM layers with dropout layers in-between, with a dropout rate of 0.2. As in the previous model, the architecture ends with two Dense layers.

\subsection{Fake news mitigation}

Having detected a source of fake news, we mitigate its spread by rating each user with a network-aware ranking score that represents how influential that user is. This ranking score weighs in the following network-based goals: firstly, we aim to immunize nodes that are followed extensively, by followers themselves having many followers; thus, we take into account the height of the subtree spanning out from them, i.e., the length of diffusion paths. Secondly, we wish to immunize nodes that spread information to multiple nodes; therefore, we take into account the size of the spanning subtree, i.e., the number of nodes. Lastly, we intend to immunize nodes that spread information faster than others; ergo, we immunize nodes having high information diffusion speed. A high ranking score by these criteria implies that a user has a high impact on news spread. After calculating this score for each user, we select the top-$k$ most harmful users to be proposed for removal or added to a blacklist.

\begin{figure}[!ht]
\centering
\includegraphics[width=1\columnwidth]{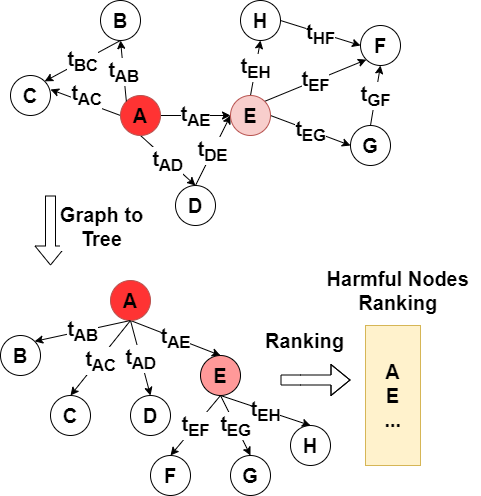}
\caption{From graph to the harmful nodes ranking}
\label{fig:net-graph}
\end{figure}

Figure~\ref{fig:net-graph} presents an example that evaluates every node that may spread a harmful article for a given source node. Firstly, given a weighed directed graph $G=(V, E)$ with positive costs $t_{u,v} \in \tau$, where $\tau = \{t_{u,v} |  (u,v,t_{u,v}) \in E \}$, with each $t_{u,v}$ measuring the time required to propagate the harmful content from $u$ to $v$, we build a minimum-cost weighted directed spanning tree (MCWDST) $T$ rooted at the detected source node $r$, using Algorithm~\ref{mcwdst}.

\begin{algorithm}[!ht]
\SetAlgoNoLine
\DontPrintSemicolon

\SetKwInOut{Input}{Input}
\SetKwInOut{Output}{Output}
\Input{the weighted directed graph $G = (V, E)$ \newline 
        the starting node $r$}
\Output{the MCWDST $T = (V_t, E_t)$ }
\emph{$V_t \gets \{ r \}$}\;\label{line1}
\emph{$E_t \gets \emptyset$}\;\label{line2}
\emph{$V_c \gets \{ r \}$}\;\label{line3}
\emph{$E_c \gets E \setminus \{ (u, r, t_{u,r})| (u, r, t_{u,r}) \in E\}$}\;\label{line4}

\While{$V_c \neq \emptyset$}{\label{line5}
    \emph{$V_n \gets \emptyset$}\;\label{line6}
    \emph{$E_n \gets \emptyset$}\;\label{line7}
    \ForEach{$n \in V_c$}{\label{line8}
        \ForEach{$(u, v, t_{u,v}) \in E_c$}{\label{line9}
            \If{$ n = u \wedge v \notin V_t $}{\label{line10}
                \emph{$V_t \gets V_t \cup \{ v \}$}\;\label{line11}
                \emph{$E_t \gets E_t \cup \{ (n, v, t_{n,v}) \}$}\;\label{line12}
                \emph{$V_n \gets V_n \cup \{ v \}$}\;\label{line13}
                \emph{$E_n \gets E_n \cup \{ (n, v, t_{n,v}) \}$}\;\label{line14}
            }
            \If{$n = u \wedge v \in V_t $}{\label{line15}
                \If{$t_{r,v} > t_{r,n} + t_{n, v} $}{\label{line16}
                    \emph{$E_t \gets E_t \setminus \{ (u', v, t'_{u',v}) | (u', v, t'_{u',v}) \in E_t \}$}\;\label{line17}
                    \emph{$E_t \gets E_t \cup \{ (u, v, t_{u,v}) \}$}\;\label{line18}
                    \emph{$E_n \gets E_n \cup \{ (u, v, t_{u,v}) \}$}\;\label{line19}
                }
            }
        }
    }
    \emph{$V_c \gets V_n$}\;\label{line20}
    \emph{$E_c \gets E_c \setminus E_n$}\;\label{line21}
}
\Return{$T = \{V_t, E_t\}$}\;\label{line22}
\caption{MCWDST for starting node $r$}
\label{mcwdst}
\end{algorithm}\DecMargin{1em}

Algorithm~\ref{mcwdst} works as follows.
We initialize 
(1) the tree's set of vertices $V_t$ to $\{r\}$ and $E_t$ to the empty set, and 
(2) the sets of untested vertices $V_c$ to $\{r\}$ and the set of untested edges $E_c$ to all edges except the ones that point to $r$ (Lines~\ref{line1}-\ref{line4}).
While there are still untested vertices (Line~\ref{line5}), we initialize, in each iteration, a set for vertices $V_n$ and one for edges $E_n$ (Lines~\ref{line6}-\ref{line7}) which are needed to update $V_c$ and $E_c$ at the end of the iteration (Lines~\ref{line20}-\ref{line21}). We pass over all vertices $n \in V_c$ and find the edges in $E_c$ where $n$ is the parent of $v$ (Lines~\ref{line8}-\ref{line9}). If the node $v$ is not present in $V_t$ then we add $v$ to $V_t$ and $V_n$ and the edge $(n, v)$ to $E_t$ and $E_n$ (Lines~\ref{line10}-\ref{line14}), otherwise, it means that there is an edge pointing to $v$ in $E_t$ (Lines~\ref{line15}-\ref{line17}). In this case, we verify if it is cheaper to pass through this edge than to use the existing one (Line~\ref{line16}). If it is cheaper, we update $E_t$ by removing the existing edge (Line~\ref{line17}) and adding the new edge to $E_n$ (Line~\ref{line18}). We update~$V_c$ with the newly added nodes in~$V_n$, and~$E_c$ by removing the nodes in~$E_n$ (Lines~\ref{line20}-\ref{line21}). When the algorithm exits the while block, it returns the propagation tree $T$. The complexity is $O(|V||E|)$. Having obtained this tree, we evaluate the potential harmfulness of each node using a ranking function and sort the nodes in descending order of their scores to obtain a leaderboard of the most harmful nodes.

Our ranking function assesses each node $n$ (Equation~\eqref{eq:rank}) in real-time in terms of three components that meet the criteria enumerated previously:
\begin{itemize}
    \item[(1)] $H(n)$: the normalized height of the subtree of $n$ to promote nodes that have a long chain of perpetuating followers,
    \item[(2)] $A(n)$: the normalized size of the subtree of $n$ to promote nodes that reach many other nodes, and
    \item[(3)] $f_{t}(n)$: a function applied to the timestamps of the descendants of $n$ to promote nodes of high information diffusion speed.
\end{itemize}

\begin{equation}\label{eq:rank}
rank(n) = H(n) + A(n) +  (1 - f_{t}(n))
\end{equation}

The normalized weight $H(n) = \frac{h_{max\_n}}{h_{max\_\mathcal{R}}}$, where $h_{max\_n}$ is the maximum height of the subtree of $n$ and $h_{max\_\mathcal{R}}$ the maximum height of the whole tree, takes into account the depth of the propagation path of harmful content, i.e., how many levels in the graph can be affected by harmful content spread by node $n$. This factor is in range $[0, 1]$ as $0 \leq h_{max\_n} \leq h_{max\_\mathcal{R}}$.
Note, for leaf nodes, $H(n)$ is always 0 as a leaf does not have a subtree.

The normalized area $A(n) = \frac{A_n}{A_\mathcal{R}}$, where $A_n$ is the area of the subtree of $n$ and $A_\mathcal{R}$ is the area of the whole tree, encodes into the ranking function the hole propagation paths, i.e., how many nodes are affected by node $n$. This factor is in range $(0, 1]$ as $0 < A_n \leq A_\mathcal{R}$.
Note, for leaf nodes, $A(n)$ is always 0 as a leaf does not have a subtree.

We include $f_t(n)$ to weigh the normalized timestamp of spreading fake news (the cost of each edge). For a node $n$ with $k$ edges and a sorted set of timestamps (costs of edges) $\tau_n = \{ t_i | t_i = t_{n,v} \wedge (n,v,t_{n,v}) \in E \wedge i = \overline{1, k} \}$, we normalize the value of a timestamp $t_i$ in $[0,1]$ using Equation~\eqref{eq:minmax}.
Note: this function is applied only to nodes that have descendants in the tree.
For leaf nodes, we do not apply the function as they have no descendants. Thus, in the ranking function $rank(n)$, the value of $f_t(n)$ for leaf nodes is always 0.  

\begin{equation}\label{eq:minmax}
    t_i'=
    \begin{cases}
        \frac{t-\min_{t_j\in \tau_n}{(t_j)}}{\max_{t_j\in \tau_n}{(t_j)}-\min_{t_j\in \tau_n}{(t_j)}} & {\scriptstyle \min_{t_j\in \tau_n}{(t_j)} \neq \max_{t_j\in \tau_n}{(t_j)}}  \\
        0.5 & {\scriptstyle \min_{t_j\in \tau_n}{(t_j)} = \max_{t_j\in \tau_n}{(t_j)}}
    \end{cases}
\end{equation}

We consider three ways of computing $f_t(n)$:
\begin{itemize}
    \item[(1)] average timestamp value, $average(t_n)= \frac{1}{k}\sum_{i=0}^k t_i$, which weighs in the average propagation time;
    \item[(2)] median timestamp (Equation~\eqref{eq:median} using the sorted order), which may be more descriptive than the average when there are outliers in the sequence that might skew the average value; and
    \item[(3)] the ratio of minimum to maximum timestamp value, $ratio(t_n) = \frac{{min}_{t_i\in \tau_n}(t_i)} {{max}_{t_i \in \tau_n}(t_i)}$, which emphasizes extremes. 
\end{itemize}

If a node has no descendants, the value of the function $f_t(n)$ is 0. Regardless of the computational approach for $f_t(n)$, the function is in the range $[0,1]$.

\begin{equation}\label{eq:median}
    median({t_n}) = 
    \begin{cases} 
        t_{\frac{k}{2}} & \text{if } k \text{ is even} \\
        \frac{t_{\frac{k-1}{2}} +  t_{\frac{k+1}{2}}}{2} & \text{if } k \text{ is odd}
    \end{cases}
\end{equation}

In effect, $f_t(n)$ measures the latency by which a node $n$ can disseminate fake news to its descendants if such exists; for a harmful node, this value is close to 0. To express speed, we use $1-f_t(n)$ in the ranking function. Given that $rank(n) \in (1,\,3]$, a harmful node will have a value that is close to 3. After scoring all nodes, we sort them in descending order of their scores and choose the top-$k$ nodes, which consequently have a high potential to infect the network with fake news.
Note, $rank(n)=1$ for a leaf node as $A(n)=0$, $H(n)=0$, and $1-f_t(n)=1$

Returning to Figure \ref{fig:net-graph}, node $A$ is marked by the detection module as a source node that spreads fake news. We build a propagation tree for $A$ using Algorithm~\ref{mcwdst}, score each node in that tree using the ranking function, and create a list of potentially harmful nodes. In effect, we find $E$, which may spread content from $A$ to its followers, to be the second most harmful node after $A$.

After extracting node ranks, we start immunizing the network from the root node to the leaves. To do so, we suggest a minimally interventionist approach of lowering the rank of a harmful post to the end of its followers' feed. Thus, the harmful content will require more time to reach other users and followers if they are not expressly looking for it by browsing the feed. To mitigate the behavior where users search for content, we can monitor the next ranked nodes to stop the network infection and apply the same strategy as for the root. Such immunization is real-time, allowing for continuous monitoring nodes that may reshare the harmful content.

\section{Implementation }\label{sec:implementation}

In this section, we present the details of our implementation in Python 3. The code is publicly available on GitHub at \url{https://github.com/DS4AI-UPB/MCWDST}.

\subsection{Preprocessing}

We load the data using \href{https://pandas.pydata.org/}{\textit{pandas}}~\cite{mckinney-proc-scipy-2010}. For processing, we use: 
(1) \href{https://scikit-learn.org/stable/}{\textit{SciKit-Learn}}~\cite{scikit-learn} \textit{LabelEncoder} to transform labels from categorical to numerical, 
(2) regular expressions to clean the text, 
(3) \href{https://www.nltk.org/}{\textit{NLTK}}~\cite{Bird2009} \textit{WordNetLemmatizer} to extract the lemmas, and
(4) \href{https://keras.io/}{\textit{Keras}}~\cite{Chollet2015keras} \textit{Tokenizer} to vectorize the clean text corpus.

To obtain word embeddings, we consider two approaches:
(1) training a Word2vec CBOW model on each dataset using \href{https://radimrehurek.com/gensim/}{\textit{Gensim}}~\cite{rehurek2011gensim}, and
(2) using the pre-trained models \textit{GloVe pre-trained} and \textit{Word2Vec pre-trained}.

We train 100-dimensional \textit{Word2Vec CBOW}~\cite{Mikolov2013} embeddings using a learning rate of 0.025 and a word window size of 5 for 5 epochs. We chose CBOW over Skip-Gram because it is faster to train and it uses distributed representations of context to predict words instead of using a distributed representation of the words to predict the context. For the pre-trained embeddings we chose: 
(1) \textit{GloVe pre-trained}: a 100-dimensional GloVe~\cite{pennington-etal-2014-glove} embedding trained on Wikipedia 2014 and Gigaword 5, and 
(2) \textit{Word2Vec pre-trained}: 300-dimension Word2Vec embeddings trained on Wikipedia data.

\subsection{Fake news detection}

In the Detection module, we use the  \href{https://keras.io/}{\textit{Keras}} interface of the \href{https://www.tensorflow.org/}{\textit{TensorFlow}}~\cite{tensorflow2015}
library to build our networks; we compile the model using the Adam optimizer with binary cross-entropy as the loss function, train the models for 200 epochs using a validation split of 0.2 and a batch size of 128, and add an early-stopping callback that monitors the validation loss.

\subsection{Fake news mitigation}

We construct propagation trees by reading the graph's adjacency matrix, along with the timestamps associated with each node. The propagation tree is the minimum-cost weighted directed spanning tree starting from a source node that spreads fake news. We iterate through all nodes in the tree and assign a ranking score to each. To calculate the function of descendant timestamps, we use the built-in Python library \textit{statistics}. We sort the scores dictionary by descending values and take the first $k$ nodes as the most harmful. We visualize the tree by
\href{https://graphviz.org/}{\textit{Graphviz}} \cite{Gansner2000graphviz}.

\subsection{Graphical user interface}

\begin{figure}[!htbp]
\centering
\includegraphics[width=0.78\columnwidth]{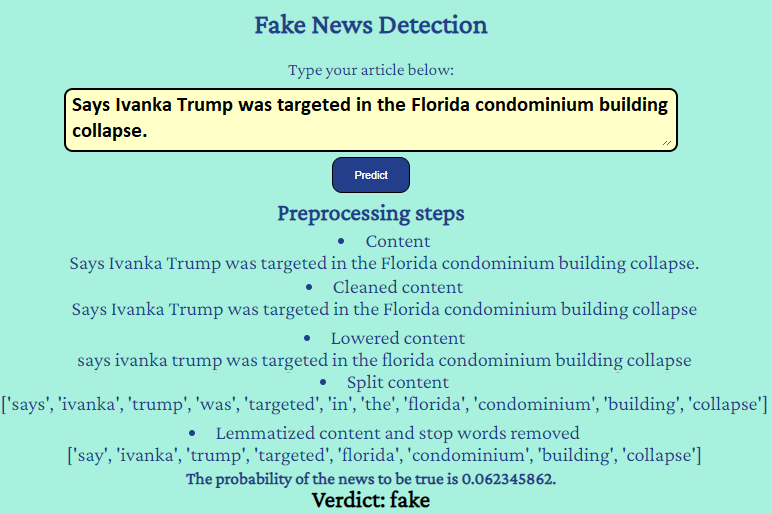}
\caption{Fake news detection page}
\label{fig:det-gui}
\end{figure}

\begin{figure}[!htbp]
\centering
\includegraphics[width=0.78\columnwidth]{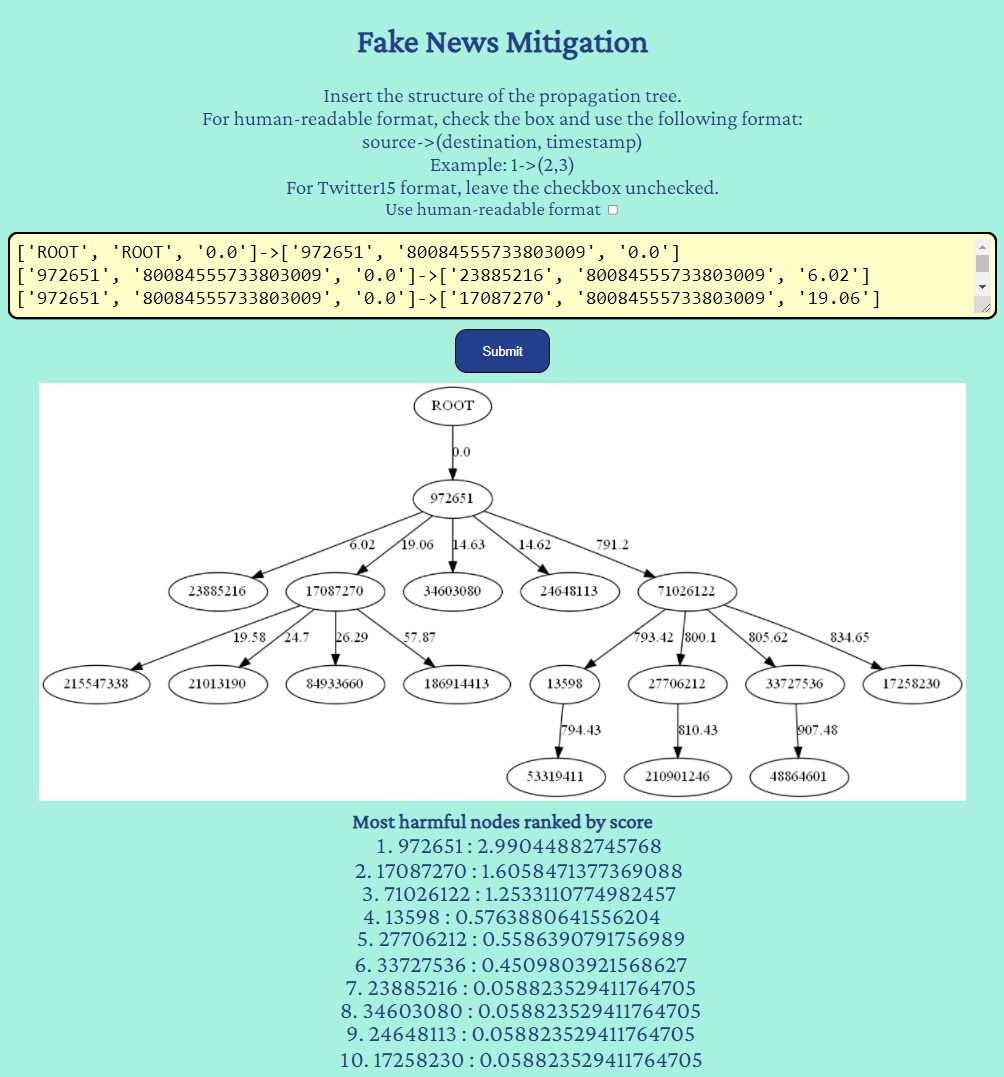}
\caption{Fake news mitigation page}\label{fig:miti-gui}
\end{figure}

We build a user-friendly graphical user interface consisting of three pages: a landing page, a fake news classification page, and a fake news mitigation page. For detection (Figure~\ref{fig:det-gui}), the user inputs the content of the article or post in a text box, clicks the ``Predict'' button, and gets the classification result as a verdict; the results of preprocessing are thereby applied to the text, with stop words and punctuation removed, and lemmatization of the remaining words. Figure~\ref{fig:miti-gui} presents the fake news mitigation page. The user can opt to visualize a node's minimum-cost weighted directed spanning tree in two formats: human-friendly format as source --- (destination, timestamp), or Twitter15~\cite{liu2015real} format. The format is toggled by a check box. Upon pressing ``Submit'', the tree image is shown along with a ranking of the most harmful nodes therein according to our evaluation metric.

\section{Experimental Results}\label{sec:results}

In this section, we present the datasets we use for evaluation, compare the performance of the two proposed architectures, and analyze how each metric in our ranking function performs, individually and collectively, in mitigating harmful content propagation.

\subsection{Datasets and Preprocessing Analysis}

We use four datasets to evaluate the detection models and one dataset to evaluate the mitigation metrics. We split each dataset for classification in training and testing sets using an 80\%-20\% ratio. From the training set, we extract 20\% for validation.

The \href{https://github.com/several27/FakeNewsCorpus}{\textit{Fake News Corpus (FNC)}}
dataset contains over 9,4 million articles labeled using 11 predefined labels. We chose a balanced subset of 20\,000 articles tagged as either reliable or fake. As the original dataset is automatically filtered, we manually verified each article's content to ensure it is correctly labeled.

The \href{https://www.kaggle.com/c/fake-news/}{\textit{Kaggle}} 
dataset, downloaded from the Kaggle Fake News Detection challenge, consists of 20\,800 articles tagged as reliable or unreliable. After eliminating null rows, the resulting dataset consists of 10\,387 reliable and 10\,374 unreliable tagged news.

The \textit{GossipCop}~\cite{shu2020fakenewsnet} dataset consists of 22\,155 news titles, article' URLs, tweet IDs of users that retweeted the article, and tags that label articles as true or fake. As the dataset is initially imbalanced, we select 5\,323 articles representing both tags.

The \textit{Fakeddit}~\cite{Nakamura2020Fakeddit} is a multimodal dataset that allows 2-way, 3-way, or 6-way classification using images, text, or both. We use the title of the post and the 2-way label to perform our task. We edit the dataset to exclude entries that contain images and select 107\,742 posts with a 1:1 true-to-fake ratio.

The \textit{Twitter15}~\cite{liu2015real} contains news articles and their labels, along with information on the directed network structure among users, followers, and followees. We use this dataset to analyze the performances of our mitigation module. It consists of 1\,490 news articles labeled as true, unverified, non-rumor, or false. We select 361 articles identified as false, construct their propagation traces, and compute scores for the nodes.

Table~\ref{datasets-content} presents the features of the datasets with cleaned content, where every article contains words that are lemmatized. In terms of average and maximum text length, we can group the datasets into two categories: large ones (i.e., FNC, Kaggle) and small ones (i.e., Fakeddit, GossipCop). The articles in FNC and Kaggle consist of sequences of sentences or phrases, while the posts in Fakeddit and GossipCop consist of a summary sentence. In terms of size, the largest dataset is Fakeddit and the smallest is GossipCop.

\begin{table}[!htbp]
\caption{Post-processing stage datasets details}\label{datasets-content}
\centering
\resizebox{\columnwidth}{!}{
\begin{tabular}{|l|r|r|r|r|r|}
\hline
\textbf{Dataset} & \begin{tabular}[c]{@{}c@{}}\textbf{Number of}\\\textbf{Article}\end{tabular} & \begin{tabular}[c]{@{}c@{}}\textbf{Average}\\\textbf{Tokens}\end{tabular} & \begin{tabular}[c]{@{}c@{}}\textbf{Maximum}\\\textbf{Tokens}\end{tabular} & \begin{tabular}[c]{@{}l@{}}\textbf{True}\\\textbf{Articles}\end{tabular}& \begin{tabular}[c]{@{}c@{}}\textbf{Fake}\\\textbf{Articles}\end{tabular}\\ \hline
FNC       &  20\,000 & 412 & 10\,736 & 10\,000 & 10\,000 \\ \hline
Kaggle    &  20\,761 & 419 & 12\,059 & 10\,387 & 10\,374 \\ \hline
GossipCop &  10\,646 &   8 &      26 &  5\,323 &  5\,323 \\ \hline
Fakeddit  & 107\,742 &   6 &      41 & 53\,975 & 53\,767 \\ \hline
\end{tabular}
}
\end{table}

\begin{figure}[!ht]
    \centering
    \captionsetup{justification=centering}
    \subfloat[3BiLSTN with FNC]{%
        \includegraphics[width=\columnwidth]{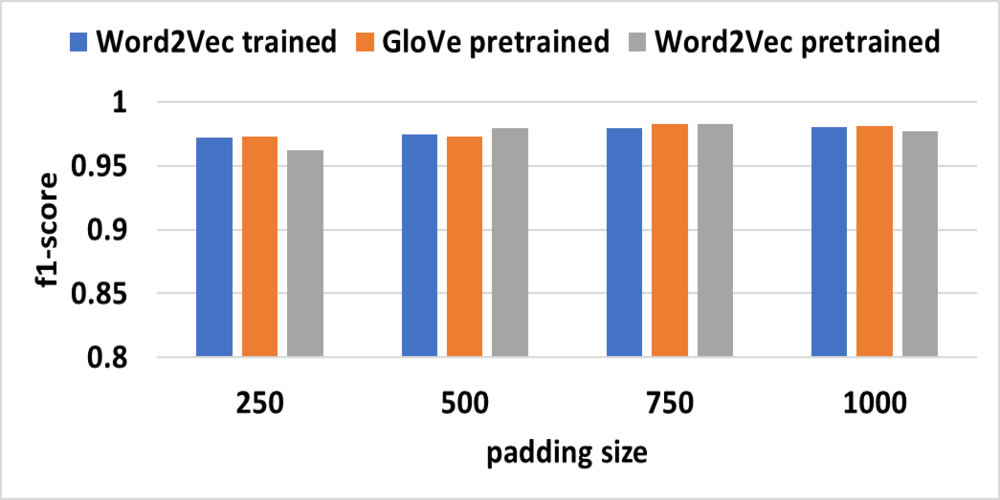}%
        \label{fig:f1-stacked1}%
        }%
    \hfill%
     \subfloat[3BiLSTN with Kaggle]{%
        \includegraphics[width=\columnwidth]{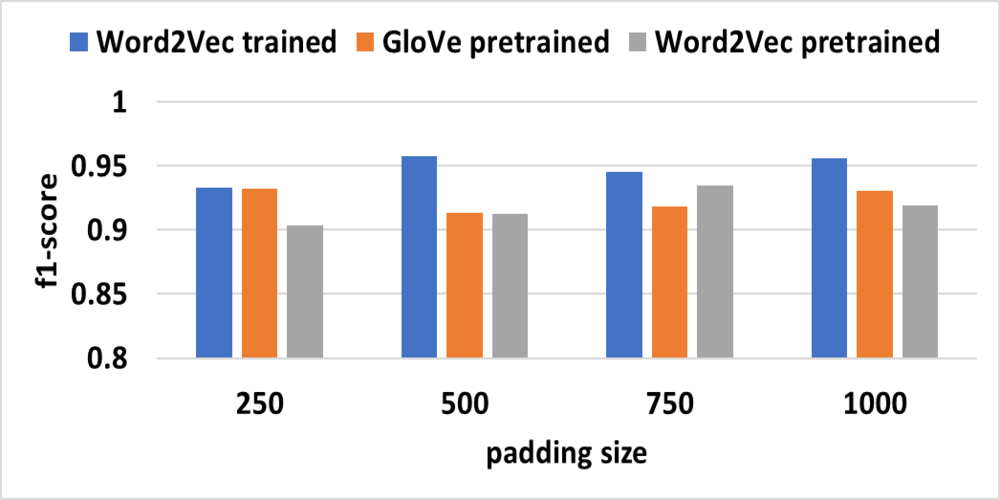}%
        \label{fig:f1-stacked2}%
        }%
    \hfill%
    \subfloat[CNN-3BiLSTN with FNC]{%
        \includegraphics[width=\columnwidth]{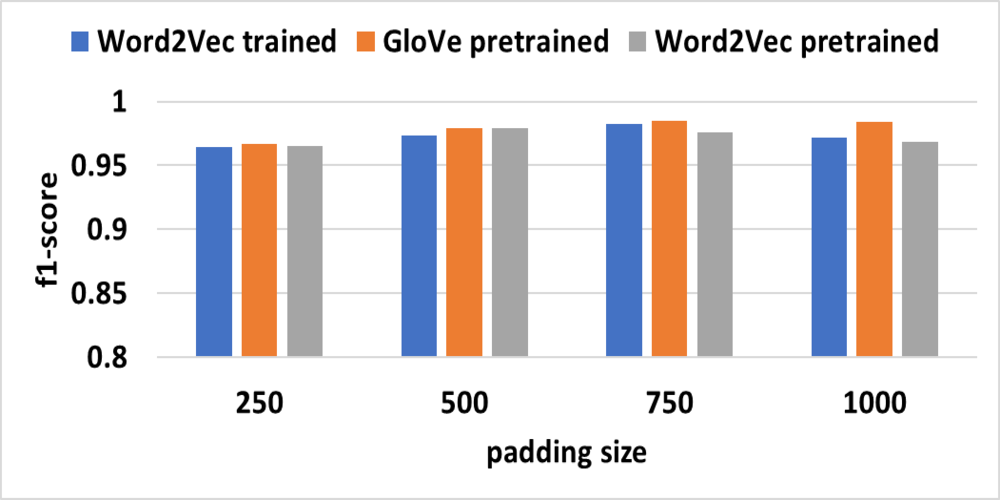}%
        \label{fig:f1-conv1}%
        }%
    \hfill%
    \subfloat[CNN-3BiLSTN with Kaggle]{%
        \includegraphics[width=\columnwidth]{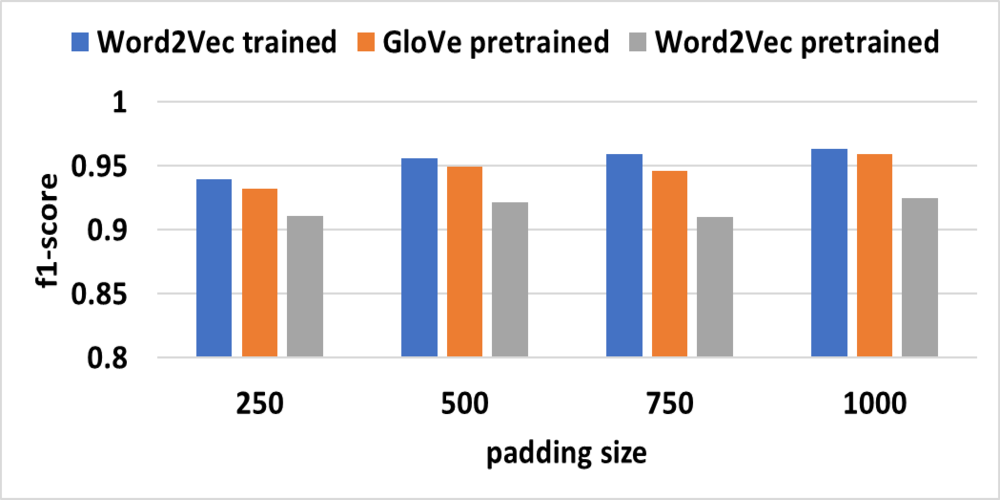}%
        \label{fig:f1-conv2}%
        }%
    \caption{The influence of padding on the f1-scores by architecture}
    \label{f1-scores-fig}
\end{figure}

\begin{table*}[!htbp]
\centering
\caption{Classification results (Notes: 1. results are the average of 10 runs and 2. \textbf{bold} marks the overall best result)}
\label{tab:class_res}

\begin{tabular}{|l|c|c|r|r|r|r|}
\hline
\textbf{Dataset} & \textbf{Embedding} & \textbf{Model} & \textbf{Accuracy} & \textbf{Precision} & \textbf{Recall} & \textbf{F1-Score} \\ \hline \hline
\multirow{6}{*}{FNC} & \multirow{2}{*}{\begin{tabular}[c]{@{}c@{}}Word2Vec\\ trained\end{tabular}} & CNN-3BiLSTM & 97.15    & 97.17     & 97.13  & 97.14  \\ \cline{3-7} 
 &  & 3BiLSTM & 98.02    & 98.02     & 98.02  & 98.02  \\ \cline{2-7} 
 & \multirow{2}{*}{\begin{tabular}[c]{@{}c@{}}GloVe\\ pre-trained\end{tabular}} & CNN-3BiLSTM & \textbf{98.37}    & 98.44     & 98.34  & 98.37  \\ \cline{3-7} 
 &  & 3BiLSTM & 98.15    & 98.16     & 98.13  & 98.14  \\ \cline{2-7} 
 & \multirow{2}{*}{\begin{tabular}[c]{@{}c@{}}Word2Vec\\ pre-trained\end{tabular}} & CNN-3BiLSTM & 96.85    & 96.87     & 96.84  & 96.84  \\ \cline{3-7} 
 &  & 3BiLSTM & 97.67    & 97.69     & 97.66  & 97.67  \\ \hline \hline

\multirow{6}{*}{Kaggle} & \multirow{2}{*}{\begin{tabular}[c]{@{}c@{}}Word2Vec\\ trained\end{tabular}} & CNN-3BiLSTM & \textbf{96.38}    & 96.39     & 96.38  & 96.38  \\ \cline{3-7} 
 &  & 3BiLSTM & 95.64    & 95.70     & 95.65  & 95.64  \\ \cline{2-7} 
 & \multirow{2}{*}{\begin{tabular}[c]{@{}c@{}}GloVe\\ pre-trained\end{tabular}} & CNN-3BiLSTM & 95.95    & 95.95     & 95.95  & 95.95  \\ \cline{3-7} 
 &  & 3BiLSTM & 93.06    & 93.09     & 93.07  & 93.06  \\ \cline{2-7} 
 & \multirow{2}{*}{\begin{tabular}[c]{@{}c@{}}Word2Vec\\ pre-trained\end{tabular}} & CNN-3BiLSTM & 92.46    & 92.46     & 92.46  & 92.46  \\ \cline{3-7} 
 &  & 3BiLSTM & 91.93    & 91.93     & 91.93  & 91.93  \\ \hline \hline

\multirow{6}{*}{GossipCop} & \multirow{2}{*}{\begin{tabular}[c]{@{}c@{}}Word2Vec\\ trained\end{tabular}} & CNN-3BiLSTM & 68.82    & 71.28     & 69.41  & 68.28  \\ \cline{3-7} 
 &  & 3BiLSTM & 69.24    & 70.45     & 69.55  & 68.98  \\ \cline{2-7} 
 & \multirow{2}{*}{\begin{tabular}[c]{@{}c@{}}GloVe\\ pre-trained\end{tabular}} & CNN-3BiLSTM & 73.66    & 73.69     & 73.60  & 73.60  \\ \cline{3-7} 
 &  & 3BiLSTM & 74.08    & 74.08     & 74.03  & 74.04  \\ \cline{2-7} 
 & \multirow{2}{*}{\begin{tabular}[c]{@{}c@{}}Word2Vec\\ pre-trained\end{tabular}} & CNN-3BiLSTM & \textbf{74.97}    & 75.26     & 75.03  & 74.92  \\ \cline{3-7} 
 &  & 3BiLSTM & 73.61    & 74.32     & 73.72  & 73.47  \\ \hline  \hline
 
\multirow{6}{*}{Fakeddit} & \multirow{2}{*}{\begin{tabular}[c]{@{}c@{}}Word2Vec\\ trained\end{tabular}} & CNN-3BiLSTM & 71.38    & 71.59     & 71.40  & 71.32  \\ \cline{3-7} 
 &  & 3BiLSTM & 71.91    & 71.97     & 71.88  & 71.87  \\ \cline{2-7} 
 & \multirow{2}{*}{\begin{tabular}[c]{@{}c@{}}GloVe\\ pre-trained\end{tabular}} & CNN-3BiLSTM & 74.26    & 74.27     & 74.25  & 74.25  \\ \cline{3-7} 
 &  & 3BiLSTM & 75.66    & 75.89     & 75.70  & 75.62  \\ \cline{2-7} 
 & \multirow{2}{*}{\begin{tabular}[c]{@{}c@{}}Word2Vec\\ pre-trained\end{tabular}} & \multicolumn{1}{l|}{CNN-3BiLSTM} & 73.80    & 73.90     & 73.78  & 73.76  \\ \cline{3-7} 
 &  & \multicolumn{1}{c|}{3BiLSTM} & \textbf{76.90}    & 77.27     & 76.89  & 76.82  \\ \hline
\end{tabular}
\end{table*}

\subsection{Fake news detection results}

Table~\ref{tab:class_res} provides the classification results of the proposed architectures using the 3 embedding methods, i.e., Word2Vec trained, GloVe pre-trained, and Word2Vec pre-trained. The results are the average for each score after 10 distinct executions using random seeding. We use a default padding size of 10 for small-length data (Fakeddit and GossipCop) and 1\,000 for large-length (i.e., Kaggle and FNC) data. CNN-3BiLSTM obtains the overall best results on Fakeddit, Kaggle, and FNC, although the difference between CNN-3BiLSTM and 3BiLSTM is relatively small. On FNC, the best accuracy is obtained when using CNN-3BiLSTM with GloVe pre-trained, i.e., 98.37. The difference between CNN-3BiLSTM and BiLSTM with the same embedding is insignificant, i.e., 0.22. CNN-3BiLSTM with Word2Vec trained registers an accuracy of 96.38 on Kaggle, with only a small increase of 0.74 over BiLSTM with the same embedding. On GossipCop, CNN-3BiLSTM with Word2Vec pre-trained obtains the best accuracy, i.e., 74.97. On Fakeddit, 3BiLSTM with Word2Vec pre-trained embeddings obtains the best accuracy. We emphasize that, in fake news detection, recall is an important metric as it tells how many fake articles are correctly classified. On datasets having a large average text length (i.e., FNC and Kaggle), recall scores are between 90 and 100. On small-length data, recall is lower than 77.

Table~\ref{tab:res_comp} presents a comparison between the results obtained by our models and the current state-of-the-art models.
We can observe that the proposed architecture outperforms or obtains similar results as the current models.
MisRoB{\AE}RTa~\cite{Truica2022} and the proposed CNN-3BiLSTM with Word2Vec-trained embeddings obtain very similar results. 
The small difference between the models is a direct result of the embedding: 
MisRoB{\AE}RTa uses two context-aware stat-of-the-art transformers, i.e., BART~\cite{Lewis2020} and RoBERTa~\cite{Liu2019}, as embeddings, while we use a static embedding.

\begin{table}[!htbp]
\centering
\caption{Comparison with state-of-the-art models (Note: \textbf{bold} text marks the overall best result)}
\label{tab:res_comp}
\resizebox{\columnwidth}{!}{
\begin{tabular}{|l|l|r|r|r|r|}
\hline
\textbf{Dataset} & \textbf{Model} & \textbf{Accuracy} & \textbf{Precision} & \textbf{Recall} & \textbf{F1-Score} \\ \hline \hline

\multirow{3}{*}{Kaggle}    & \begin{tabular}[l]{@{}l@{}}CNN-3BiLSTM + \\ Word2Vec trained\end{tabular}          & \textbf{96.38} & 96.39   & 96.38  & 96.38 \\ \cline{2-6} 
                           & DEAP-FAKED~\cite{mayank2021deapfaked}   & 88.66          & N/A     & N/A    & 89.55 \\ \cline{2-6} 
                           & MisRoB{\AE}RTa~\cite{Truica2022}            & \textbf{97.85}          & 97.85   & 97.85  & 97.85 \\ \hline \hline

\multirow{2}{*}{GossipCop} & \begin{tabular}[l]{@{}l@{}}CNN-3BiLSTM + \\ Word2Vec pre-trained\end{tabular}       & \textbf{74.97} & 75.26   & 75.03  & 74.92 \\ \cline{2-6}
                           & CNN~\cite{shu2020fakenewsnet}           & 72.30          & N/A     & N/A    & 72.50 \\ \hline \hline

\multirow{2}{*}{Fakeddit}  & \begin{tabular}[l]{@{}l@{}}3BiLSTM + \\ Word2Vec pre-trained\end{tabular}          & \textbf{76.90}  & 77.27   & 76.89  & 76.82 \\ \cline{2-6}
                           & FND-NS~\cite{Raza2022}                  & 74.80          & 72.40   & 77.60  & 74.90 \\ \hline
\end{tabular}
}
\end{table}

We also study the impact of padding size on performance with large data, i.e., Kaggle and FNC. We do not include small-length data in our analysis, since their average text length is 6-8 and their content consists of one sentence; thus, their padding size range is small. We evaluate the two datasets in terms of f1-score for 4 padding sizes, namely 250, 500, 750, and 1\,000. Figure~\ref{f1-scores-fig} shows the f1-scores we obtain after 10 distinct executions using random seeding. On FNC with 3BiLSTM, the best-performing embedding strategy is Word2Vec pre-trained with a padding size of 750. On Kaggle with 3BiLSTM, the Word2Vec-trained embeddings obtain the best results when the padding size is 500. On FNC with CNN-3BiLSTM, GloVe pre-trained with a padding size of 750 achieves the best f1-score. The Word2Vec-trained embeddings with a 1\,000 padding size have the best score on Kaggle with CNN-3BiLSTM. We note that a larger text size does not necessarily translate into better results. We conclude that the padding size has a small impact on model performance.

\subsection{Fake news mitigation results}

To analyze the performance of our proposed algorithm in identifying harmful nodes by measuring the distribution of news by users, we use 361 trees with an average of 335 nodes per tree, a minimum number of nodes of 97, a maximum number of nodes of 2\,971, an average tree height of 4, a minimum height of 2, and a maximum height of 10.

We first analyze the importance of $f_t(n)$ in ranking nodes. We divide the interval $(1,3]$ into two intervals and study the influence of the function $f_t(n)$ on the obtained subintervals. Table~\ref{k-timestamp-table} shows the percentages of the scores obtained for the evaluation metric in each range, for the three strategies applied in the $f_t(n)$ function, when collecting the top-$k$ most harmful nodes with $k \in \{1\,000, 2\,000, 3\,000\}$. A higher score represents a greater capability to spread untruthful information faster. Thus, a score in $(2, 3]$ indicates a very harmful node to be immunized first, a score in $[1, 2]$ indicates a mildly harmful node to be immunized in second priority, where nodes with a score close to 1 suggest that a node is weakly to not harmful and should be immunized in last priority. A node with a score of 1 is a lead node. 
We observe that when using $median(t_n)$, our ranking function scores with higher values more harmful nodes (i.e., nodes $\in (2, 3]$) than when using $average(t_n)$ or $ratio(t_n)$ (i.e., harmful nodes are ranked in $\in (1, 2]$).
Furthermore, as we know which nodes are harmful and which are not, $rank(n)$ identifies correctly all the harmful nodes regardless of the method used to compute $f_t(n)$.

\begin{table}[!htbp]
\centering
\caption{The distribution of the top-$k$ nodes}
\label{k-timestamp-table}
\begin{tabular}{|c|c|r|r|}
\hline
$\bm{k}$ & $\bm{f_t(n)}$ & \textbf{\% in (1,\, 2{]}} & \textbf{\% in (2,\,3{]}} \\ \hline
\multirow{3}{*}{1\,000} 
 & $average(t_n)$ & 68.51 & 31.49 \\ \cline{2-4} 
 & $median(t_n)$  & 60.10 & 39.90 \\ \cline{2-4} 
 & $ratio(t_n)$   & 89.10 & 10.90 \\ \hline 
 
\multirow{3}{*}{2\,000} 
 & $average(t_n)$ & 83.31 & 16.69 \\ \cline{2-4}
 & $median(t_n)$  & 80.05 & 19.95 \\ \cline{2-4} 
 & $ratio(t_n)$   & 94.55 &  5.45 \\ \hline 
 
\multirow{3}{*}{3\,000} 
 
 & $average(t_n)$ & 90.20 &  9.80 \\ \cline{2-4}
 & $median(t_n)$  & 86.70 & 13.30 \\ \cline{2-4}
 & $ratio(t_n)$   & 96.36 &  3.64 \\ \hline
\end{tabular}
\end{table}

\begin{figure}[!htbp]
    \centering
    \captionsetup{justification=centering}
    \subfloat[$rank(n) = H(n) + A(n) + (1-average(t_n))$]{%
        \includegraphics[width=\columnwidth]{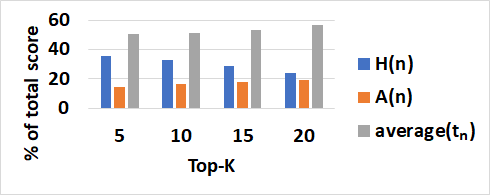}%
        \label{fig:influ-avg}%
        }%
    \hfill%
    \subfloat[$rank(n) = H(n) + A(n) + (1-median(t_n))$]{%
        \includegraphics[width=\columnwidth]{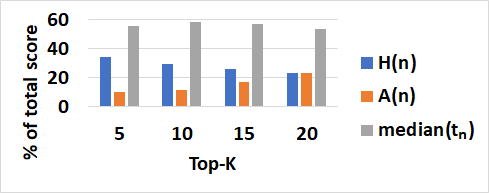}%
        \label{fig:influ-mdn}%
        }%
    \hfill%
    \subfloat[$rank(n) = H(n) + A(n) + (1-ratio(t_n))$]{%
        \includegraphics[width=\columnwidth]{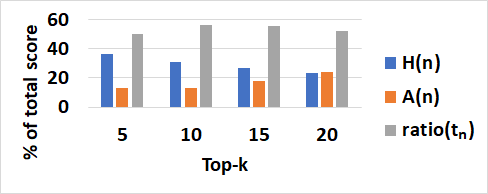}%
        \label{fig:influ-rto}%
        }%
    \caption{The influences of the 3 components of the metric in evaluating the top-$k$ most harmful nodes}
    \label{metrics-influences}
\end{figure}

Figure~\ref{metrics-influences} plots the contribution that each component of the proposed ranking function has in a node's final score. We analyze the top-$k$ most harmful nodes, where $k=\{5, 10, 15, 20\}$. The $f_t(n)$ function influences the score the most, followed by the $H(n)$ and $A(n)$ functions. Notably, the weight of $f_t(n)$ is more than 50\% in all evaluation situations. In all three subgraphs, as we increase $k$, the contribution of $H(n)$ is transferred to $A(n)$. With $k=20$, $A(n)$ surpasses $H(n)$ (Figure~\ref{fig:influ-mdn} and~\ref{fig:influ-rto}). In conclusion, the information diffusion speed has the most contribution to our ranking function. That is a reasonable design, as, for a small $k$, nodes that have long chains of followers (i.e., a longer maximum diffusion path) play a more important role than the nodes that can reach many nodes. As $k$ increases, the impact of the nodes with many followers decreases, and the impact of nodes that reach many nodes increases.

\subsection{Mitigation scalability results}

We want to test the scalability of our solution and determine the mitigation strategy performance in a real-world setting.
We collect from Twitter a graph structure containing over 10\,000 nodes and 89\,000 edges and split it into 3 subgraphs for scalability testing. 
Table~\ref{scalabilty} presents the MCWDST construction and ranking list creation time in seconds for these subgraphs.
We can observe that for a small graph, the mitigation is done almost instantly.
This is useful when we detect and target the harmful content spreaders within a disinformation network.
For a larger network, the entire process is done in near real-time, as it takes less than 5 minutes. Thus, the best strategy is to immunize the network by targeting the topmost harmful nodes.
Finally, we compare our method with two preventive network immunization strategies, i.e., NetShield~\cite{chen2015node} and SparseShield~\cite{Petrescu2021} an improved version of NetShield, and one counteractive network immunization in large networks, i.e, DAVA~\cite{Zhang2015}.

\begin{table*}[!htbp]
\centering
\caption{Mitigation scalability tests}
\label{scalabilty}
\begin{tabular}{|r|r|r|r|r|r|r|r|}
\hline
\textbf{Nodes} & \textbf{Edges} & \textbf{MCWDST (s)} & \textbf{Ranking (s)} & \textbf{DAVA} & \textbf{NetShield (s)} & \textbf{SparseShield} \\ \hline
           978 &        10\,217 &                2.01 &                 0.66 &          1.78 &                   0.29 &                  0.07 \\ \hline
        5\,210 &        49\,124 &               53.68 &                20.69 &         60.75 &                  12.02 &                  0.42 \\ \hline
       10\,210 &        89\,124 &              411.96 &               138.44 &        386.99 &                 102.13 &                  1.33 \\ \hline
\end{tabular}
\end{table*}

Although both preventive network immunization algorithms are faster, the main problems with NetShield and SparseShield that we solved with MCWDST are:
\textit{(1)} nodes are blocked in no particular order if they converge,
\textit{(2)} nodes are not ranked,
\textit{(3)} the edge weights are not used in the immunization process,
\textit{(4)} the directed graph structure is not considered, and
\textit{(5)} the number of immunized nodes is dependent on the budget.

Whereas our proposed solution gives an immunization strategy for a weighted directed graph by returning a blocking order by ranking the harmfulness of nodes.

When comparing MCWDST with DAVA, another counteractive network immunization algorithm, we observe that:
\begin{itemize}
    \item[(1)] DAVA blocks all the nodes in no particular order while our MCWDST manages to block nodes in the order given by the information diffusion paths,
    \item[(2)] DAVA does not rank nodes, while MCWDST ranks them from most to least harmful,
    \item[(3)] DAVA is dependent on the budget and immunizes a given number of nodes, while MCWDST immunizes the entire network, and
    \item[(4)] DAVA does not consider the directed graph structure, while MCWDST takes this graph property into account.
\end{itemize}

\section{Discussion}\label{sec:discussions}

This section highlights the results and lessons learned from the implementation and presents an in-depth analysis of the two modules.

\subsection{Fake news detection}

We observe that the pre-trained models (i.e., Word2Vec pre-trained and GloVe pre-trained) achieve better results on smaller datasets, i.e., GossipCop and Fakeddit, while Word2Vec trained and GloVe pre-trained obtain better results on larger datasets, i.e., FNC and Kaggle. These results of the classification models directly reflect the typical text length in each dataset. For small-length texts, it is better to use pre-trained embeddings as they better encode semantic, syntactic, and contextual information that cannot be found in short statements. This is mitigated in large datasets by the texts' length, which offers sufficient semantic, syntactic, and contextual information to train accurate and data-specific word embeddings. The proposed novel deep learning architectures offer state-of-the-art results on all tested datasets, i.e., an accuracy of 88.66 on Kaggle~\cite{mayank2021deapfaked}, 74.80 on Fakeddit~\cite{Raza2022}, and 72.30 on GossipCop~\cite{shu2020fakenewsnet}. We note that the longer texts enable the bidirectional LSTM layers with and without a CNN layer on top to detect meaningful patterns for fake news detection. Even though the experimental results show that 3BiLSTM architecture obtains good results, the CNN-3BiLSTM architecture achieves the best scores on FNC, Kaggle, and GossipCop. This result arises from the combination of a CNN layer with a MaxPooling layer, which generates new features and selects the best features among those generated. Moreover, on the Kaggle data, CNN-3BiLSTM outperforms 3BiLSTM with all embedding strategies we use.
Furthermore, we observe that, on large datasets, a large-size padding strategy does not necessarily improve performance. It follows that an ample analysis w.r.t. size and word embedding should be performed on each dataset to determine the best padding approach.

\subsection{Fake news mitigation}

The proposed ranking function takes into account 3 network-specific components for mitigating harmful nodes:
(1) $H(n)$, the length of the diffusion path,
(2) $A(n)$, the spread of information, and
(3) $f_t(n)$, the speed of information diffusion.

In our experiments on a real-world dataset, we observe that:
(1) regardless of graph size, $f_t(n)$ has the largest contribution in the function,
(2) for a small number of nodes to mitigate, the length of the diffusion path $H(n)$ adds more weight to the ranking function than the number of nodes that can be infected directly $A(n)$, and
(3) for a large number of nodes to mitigate, both $H(n)$ and $A(n)$ have a similar impact on the ranking.

As $f_t(n)$ has the largest contribution of the three components to the score value of a node, it is, in effect, the most important factor in mitigating fake news in a network, according to our empirical findings. The comparison of different calculation strategies of $f_t(n)$ indicates the impact of the potential to spread fake news quickly, while the individual evaluation of the three components of the ranking function indicates that network structure is also important for the immunization task.

\section{Conclusion}\label{sec:conclusions}

In this paper, we proposed
(1) two novel deep learning architectures for fake news detection,
(2) a real-time algorithm for building the minimum-cost weighted directed spanning tree (i.e., MCWDST), and
(3) a function that ranks harmful nodes in real-time, to combat the spread of fake news on social media.

The proposed deep learning architectures use convolutional and bidirectional LSTM layers that capture semantic, syntactic, and context information. These models offer state-of-the-art performance on the tested datasets. Our mitigation algorithm greedily constructs a minimum-cost weighted directed spanning tree for a given source node. The novel raking function takes into account network information (i.e., the length of the diffusion path, the spread of information, and the speed of information diffusion) as well as its harmful content to mitigate the spread of fake news in a network-aware, real-time manner. 
Ample benchmarking on five real-world datasets shows that the proposed models for fake news detection and the mitigation strategy are effective in stopping the spread of harmful content online.

For future research directions, we will consider using: 
(1) transformers as a way of generating word embeddings, and 
(2) graph embedding techniques to develop a more complex network-aware model that identifies harmful sources in real-time.

\section*{Acknowledgment}
The publication of this paper is supported by the National University of Science and Technology Politehnica Bucharest through the PubArt program.

\bibliographystyle{IEEEtran}
\bibliography{main}  

\vspace{-10mm}
\begin{IEEEbiography}[{\includegraphics[width=1in,height=1.25in,clip,keepaspectratio]{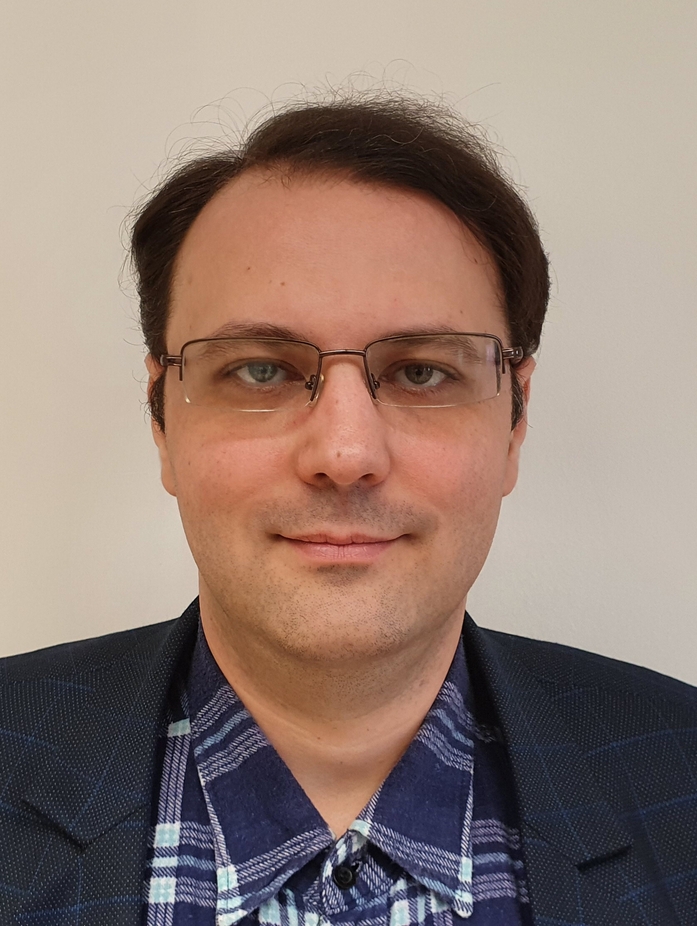}}]{Ciprian-Octavian TRUIC{\u{A}}} is an Assistant Professor of Computer Science in the Computer Science and Engineering Department, Faculty of Automatic Control and Computers, National University of Science and Technology Politehnica Bucharest, Romania. He received his Ph.D. (2018) in Data Management and Text Mining from the University Politehnica of Bucharest, Romania. He held research positions at Uppsala University (Sweden), Aarhus University (Denmark), and Lyon University (France) where he worked on topics such as Social Media Analysis, Deep Learning, Machine Learning, Graph Mining, Network Immunization, Natural Language Processing, Big Data Analytics, and Data Management.
\end{IEEEbiography}

\vspace{-10mm}
\begin{IEEEbiography}[{\includegraphics[width=1in,height=1.25in,clip,keepaspectratio]{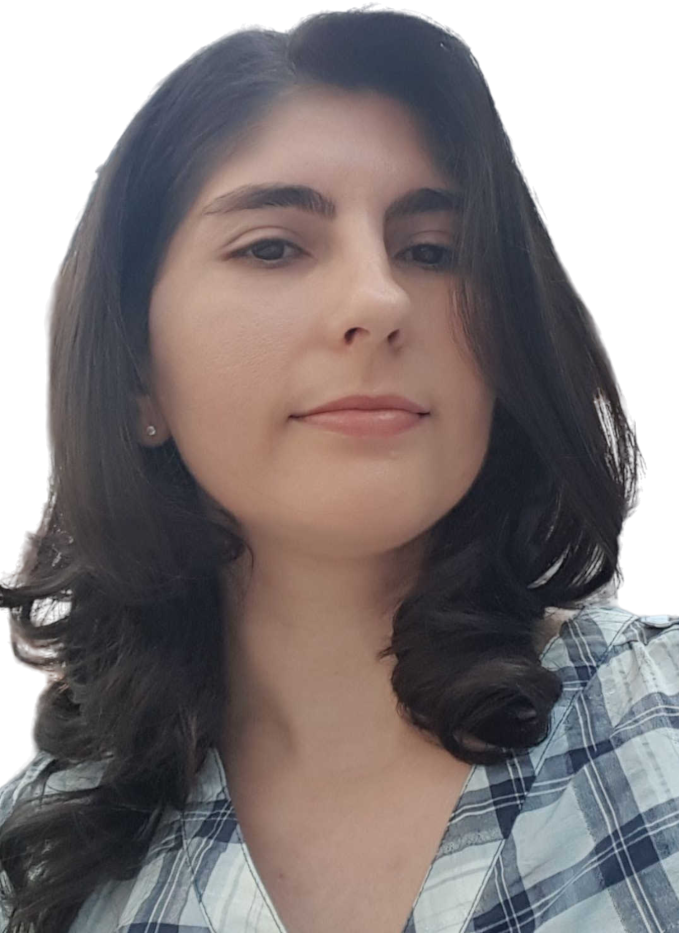}}]{Elena-Simona APOSTOL} is an Associate Professor of Computer Science in the Computer Science and Engineering Department, Faculty of Automatic Control and Computers, National University of Science and Technology Politehnica Bucharest, Romania.
She was a postdoctoral researcher at Microsoft Research Center in collaboration with INRIA Institute where she worked on state-of-the-art Big Data Analysis, Multi-Site Cloud Computing, and Bioinformatics. She also held research positions at the Department of Information Technology, Uppsala University, and Fraunhofer FOKUS Institute. Her research topics include Big Data Analytics, Data Science, Text and Data Mining, and Cloud Computing.
\end{IEEEbiography}

\vspace{-10mm}
\begin{IEEEbiography}[{\includegraphics[width=1in,height=1.25in,clip,keepaspectratio]{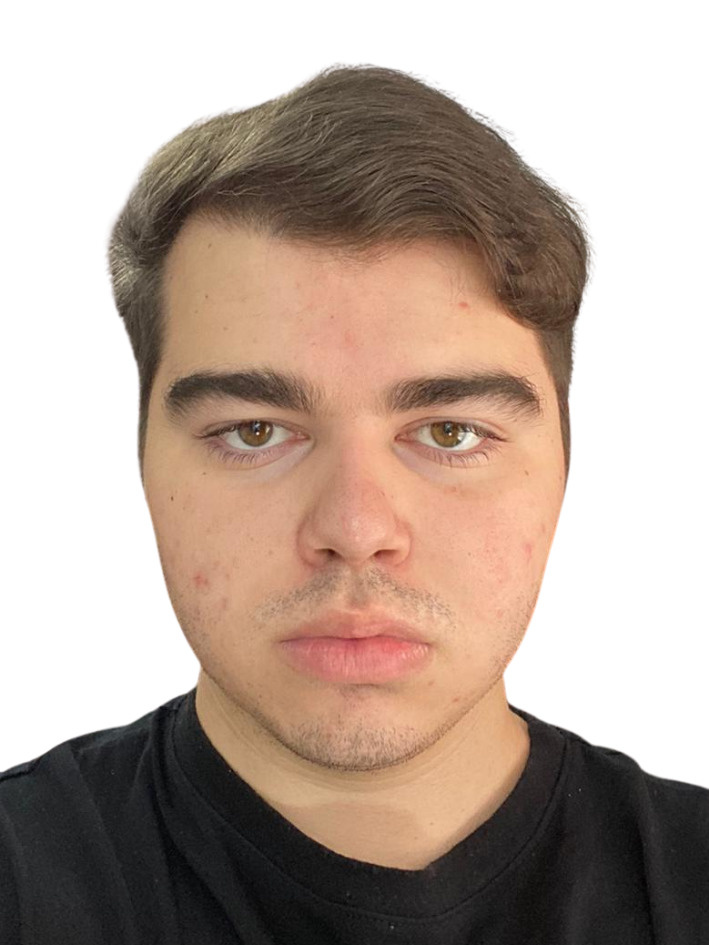}}]{Radu-C{\u{A}}t{\u{A}}lin NICOLESCU} is currently pursuing an M.Sc. in Advanced Software Services at National University of Science and Technology Politehnica Bucharest, Romania. He received a Bachelor's degree in Computer Engineering (2021) from the University Politehnica of Bucharest, Romania. His bachelor's thesis was on the subject of Fake News detection using Deep Learning methods. His fields of interest are Natural Language Processing, Computer Vision, Data Mining, and aims to find new approaches and improvements to current challenges in artificial intelligence through his research.
\end{IEEEbiography}

\vspace{-10mm}
\begin{IEEEbiography}[{\includegraphics[width=1in,height=1.25in,clip,keepaspectratio]{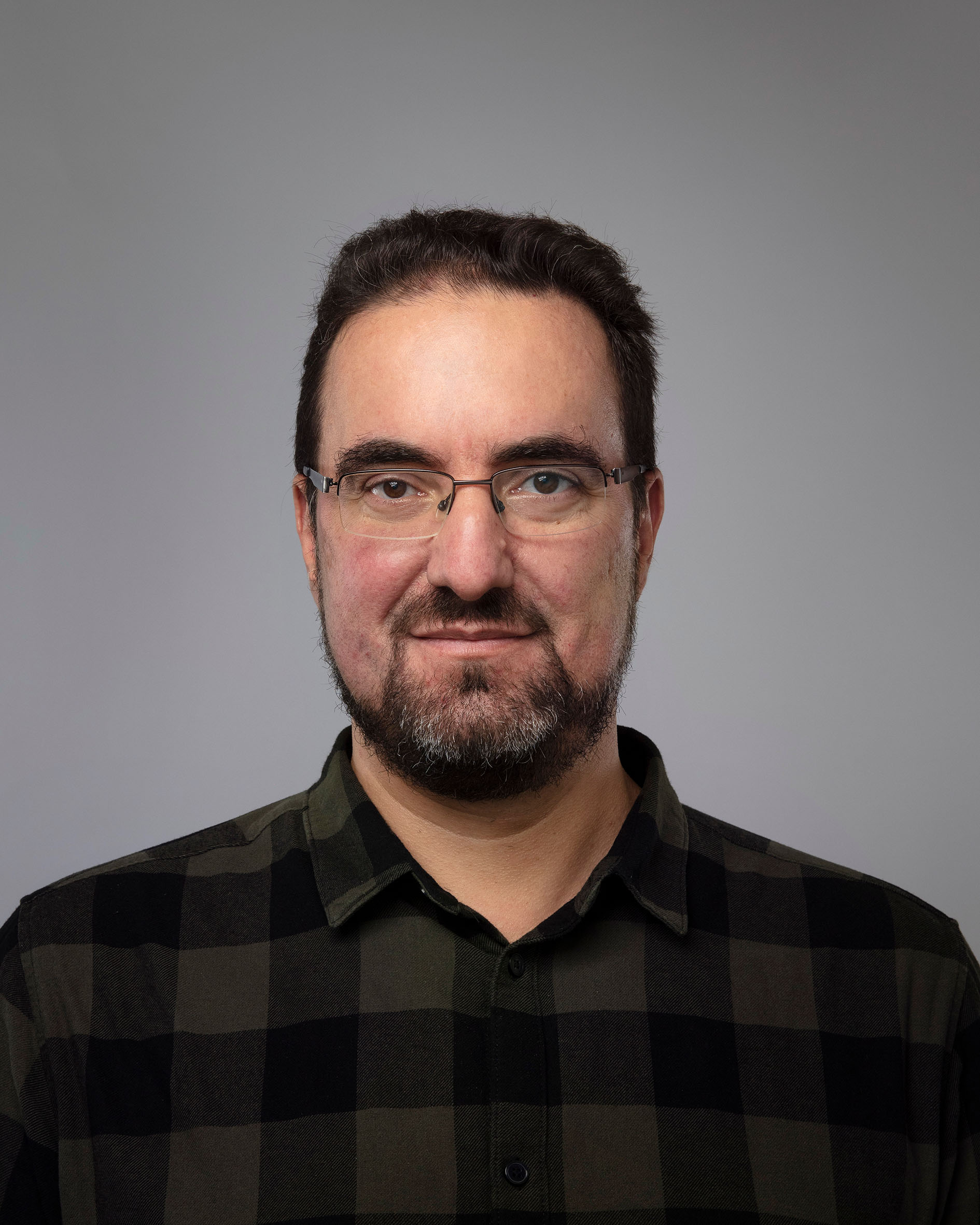}}]{Panagiotis KARRAS} is a Professor of Computer Science at the University of Copenhagen. In his research, he designs robust, scalable, and versatile methods for data access, mining, analysis, exploration, and representation. He received the PhD degree in computer science from the University of Hong Kong and the MSc degree in electrical and computer engineering from the National Technical University of Athens. He has been awarded the Hong Kong Young Scientist Award, the Singapore Lee Kuan Yew Postdoctoral Fellowship, the Rutgers Business School Teaching Excellence Fellowship, and the Skoltech Best Faculty Performance Award.
\end{IEEEbiography}

\vfill
\EOD 
\end{document}